\documentclass[fleqn,twoside]{article}
\usepackage{espcrc2}
\usepackage{graphicx}

\newcommand{\GeV}{\mathrm{GeV}}
\newcommand{\mrad}{\mathrm{mrad}}
\newcommand{\Npe}{N_\mathrm{p.e.}}
\newcommand{\sigmatrack}{\sigma_\mathrm{track}}

\begin{document}

\title{A novel type of proximity focusing RICH counter
with multiple refractive index aerogel radiator }
\author{ T.~Iijima\address[NAGOYA]{Nagoya University, Nagoya, Japan}%
\thanks{E-mail address: iijima@hepl.phys.nagoya-u.ac.jp},
S.~Korpar\address[MB]{Faculty of Chemistry and Chemical Engineering, University of Maribor, 
Slovenia}%
\address[IJS]{J.~Stefan Institute, Ljubljana, Slovenia}%
\thanks{E-mail address: samo.korpar@ijs.si},
I.~Adachi\address[KEK]{High Energy Accelerator Research Organization (KEK), Japan},
S.~Fratina\addressmark[IJS],
T.~Fukushima\address[CHIBA]{Chiba University, Japan},
A.~Gori\v sek\addressmark[IJS],
H.~Kawai\addressmark[CHIBA],
M.~Konishi\addressmark[CHIBA],
Y.~Kozakai\addressmark[NAGOYA],
P.~Kri\v zan\address[FMF]{Faculty of Mathematics and Physics, University of Ljubljana,  
Slovenia}\addressmark[IJS],
T.~Matsumoto\address[TOKYO]{Tokyo Metropolitan University},
Y.~Mazuka\addressmark[NAGOYA],
S.~Nishida\addressmark[KEK],
S.~Ogawa\address[TOHO]{Toho University, Funabashi, Japan},
S.~Ohtake\addressmark[TOHO],
R.~Pestotnik \addressmark[IJS],
S.~Saitoh\addressmark[KEK],
T.~Seki\addressmark[TOKYO],
T.~Sumiyoshi\addressmark[TOKYO],
Y.~Uchida\addressmark[KEK],
Y.~Unno\addressmark[KEK] and
S.~Yamamoto\addressmark[TOKYO]}

\begin{abstract}
A proximity focusing ring imaging Cherenkov detector, with the radiator consisting
of two or more aerogel layers of different refractive indices, has been tested in
1-4 GeV/c pion beams at KEK. Essentially, a multiple refractive index aerogel radiator allows
for an increase in Cherenkov photon yield on account of the increase in overall radiator thickness, 
while avoiding the simultaneous degradation
in single photon angular resolution associated with the increased uncertainty of the emission point.
With the refractive index of consecutive layers suitably increasing in the downstream direction,
one may achieve overlapping of the Cherenkov rings from a single charged particle.
In the opposite case of decreasing refractive index, one may obtain well separated rings.
In the former combination an approximately 40\%
increase in photon yield is accompanied with just a minor degradation in single photon angular 
resolution. The impact of this improvement on the $\pi$/K  separation at the upgraded 
Belle detector is discussed. 
\end{abstract}

\maketitle

\section{Introduction}
\label{In}

Proximity focusing ring imaging Cherenkov (RICH) counters with non-gaseous radiators
have the advantage of smaller size and are thus well suited as particle identification devices
within more complex particle spectrometers with tight space constraints. 
Such a detector with aerogel as radiator is envisaged for the upgrade~\cite{bib:superkekb-loi}
of the particle identification system of the Belle spectrometer at the KEKB 
collider~\cite{Mori:2000cg,KEKB:NIM}. In the present Belle detector,
a threshold-type Cherenkov detector (ACC)~\cite{Iijima:2000uv} enables 
a good $\pi/K$ separation. However, ACC does not provide sufficient separation for
high momentum particles around $4~\GeV/c$ in the forward end-cap region, and a proximity 
focusing RICH with aerogel as radiator is being studied as a candidate for the 
detector upgrade.

Different aerogel radiators
as well as different photon detectors have been investigated in search of an optimal counter,
which would provide the required 4-5 $\sigma$ pion-kaon separation in the 1-4 GeV/c momentum range  
\cite{Matsumoto:2003wv}.
Good velocity resolution of a charged particle track requires good resolution of the
Cherenkov angle of a single photon and many detected photons. The number of photons
may be increased by increasing the radiator thickness, but the price to pay is an increased uncertainty
of the emission point, i.e. single photon angular resolution. However, by judiciously choosing the
refractive indices of consecutive aerogel radiator layers, one may achieve overlapping
of the corresponding Cherenkov rings on the photon detector. This represents a sort of focusing
of the photons within the radiator, and eliminates or at least considerably reduces the spread
due to emission point uncertainty. Another possibility is the opposite,
i.e. to obtain separate rings from separate aerogel layers in which the emission point
uncertainty is given by the thickness of the corresponding layer. One may also try
combinations of the two main schemes. Note that such a tuning of  
 refractive index for individual layers is only possible with aerogel, 
which may be produced with any desired refractive index in the range 1.006-1.06.%??
%where the refractive index is controllable over a certain range.

While the principle was first discussed in~\cite{focus-first},  
the present work reports on measurements made with different combinations of aerogel
radiator layers.
First we present estimates of the relations between the radiator thicknesses and
refractive indices in order to obtain either overlapping of the Cherenkov rings
or their adequate separation. Next the experimental apparatus is described and the
results of measurements and analysis are given. Finally, we discuss the
possibilities of such a detector to meet the expectations for the upgrade of the
Belle particle identification system.

\begin{figure}[t]
 \begin{center}
% centerline{\includegraphics[scale=.4]{figures/ai_optimal_trl_sigev_mod.eps}}
  \centerline{\includegraphics[scale=.4]{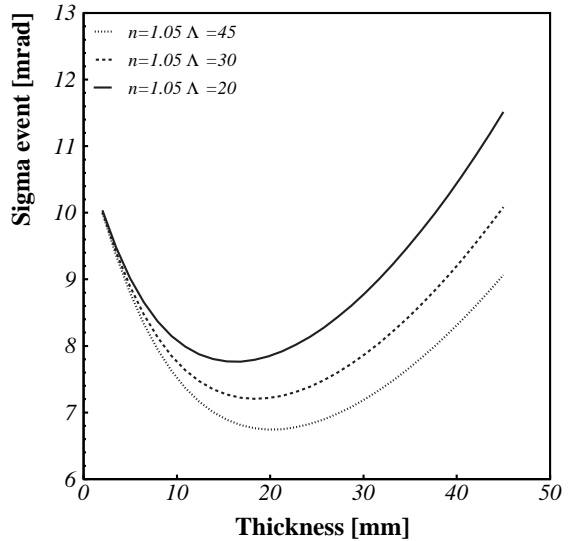}}
 \vskip -1cm
  \caption{\label{fig:optimal-d-normal}%
  The variation of the figure of merit, $\sigma_{\theta}/\sqrt{\Npe}$, with 
radiator thickness in the case of a single radiator, proximity focusing RICH,
for three different transmission lengths at 400~nm ($\Lambda$ in mm).}
 \end{center}
\end{figure}

\section{RICH with multiple refractive index radiator}

The key issue in the performance of a proximity focusing RICH counter is to
improve the Cherenkov angle resolution per track 
$\sigmatrack = \sigma_{\theta}/\sqrt{\Npe}$.
By using a thicker aerogel radiator, the average number of 
detected photons $\Npe$ can be increased,
but the single photon  Cherenkov angle resolution  $\sigma_{\theta}$ degrades 
due to the increased uncertainty in the photon emission point.
As can be seen from Fig.~\ref{fig:optimal-d-normal}, the optimal
thickness should be around $20~\mathrm{mm}$, which was also 
verified in previous investigations~\cite{Matsumoto:2003wv}.

\begin{figure}[t]
 \begin{center}
%  \centerline{\includegraphics[scale=.9]{figures/multi-radiator.eps}}
  \centerline{\includegraphics[scale=.9]{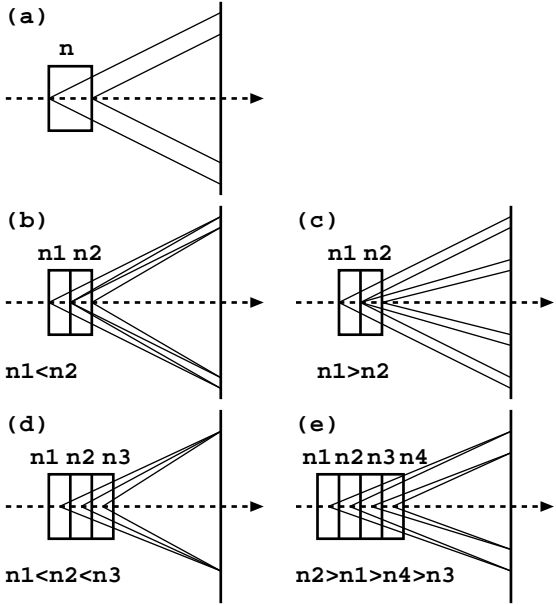}}
  \vskip -1cm
  \caption{\label{fig:multi-radiator}%
  Principle of dual radiator and multiple radiator ring imaging Cherenkov counter:
  (a) single radiator,
  (b) focusing dual radiator,
  (c) defocusing dual radiator,
  (d) focusing multiple radiator 
  and (e) defocusing multiple radiator RICH. Only photons from the middle of 
the radiator are shown in (d) and (e).
  }
 \end{center}
\end{figure}

One way to solve this problem is
to use a ``dual radiator'' scheme, where one images Cherenkov photons from
two aerogel radiators with different refractive indices
as shown in Fig.~\ref{fig:multi-radiator}.
In the first combination (Fig.~\ref{fig:multi-radiator}.b)
two aerogel radiators are used with slightly different indices,
where the one with the lower refractive index  is positioned 
upstream.
If the indices of the two radiators are well adjusted,
the corresponding two rings overlap.
In the following this combination will be referred to as ``focusing combination''.

The other possibility is a ``defocusing combination'',
in which the aerogel with higher index is positioned upstream
(Fig.~\ref{fig:multi-radiator}.c).
If the difference of the indices of the two aerogel tiles is appropriately
chosen, the
two radiators produce two well separated rings with good resolution.

A naive extension of the dual radiator combination
is to use more than two aerogel radiators (``multiple radiator'').
For the focusing combination, the indices of aerogels should
gradually increase from the upstream to the downstream layer.
Again, if the index of each layer is well chosen
(Fig.~\ref{fig:multi-radiator}.d),
the angular resolution of the ring will not be
appreciably deteriorated in spite of a thicker radiator.
The defocusing dual radiator may
be extended to four radiators
by introducing the focusing combination in every two layers
as shown in Fig.~\ref{fig:multi-radiator}.e.
In this case, the RICH has four radiators with different indices
where the first two radiators in the upstream position create a
larger ring, and the other two radiators a smaller ring.

\section{Radiator parameters and expected performance}

The relation between the refractive indices in the focusing combination
in the dual radiator case is derived as follows.
%by employing Fig.~\ref{fig:multi-radiator}.b.
Neglecting the refraction at aerogel boundaries and 
denoting the thickness of a single radiator tile as $d$, the distance 
between the center of the downstream radiator and the photon detector plane as $L$, and the
Cherenkov angles in both upstream and downstream radiator as $\theta_1$ and   
$\theta_2$ respectively, 
the condition that the 
two rings should overlap can be written as $L \tan{\theta_2} = (L+d) \tan{\theta_1}$.
From this condition the following approximate relation can be derived
$$n_2-n_1 = \frac{d}{ n_1 L} \bigl[ n_1^2-1-\frac{(mc)^2}{p^2} \bigr],$$
where $m$ and $p$ are the pion mass and momentum.

\begin{figure}
 \begin{center}
%  \centerline{\includegraphics[scale=.5]{figures/ai_sigth_p_mul.eps}}
  \centerline{\includegraphics[scale=.5]{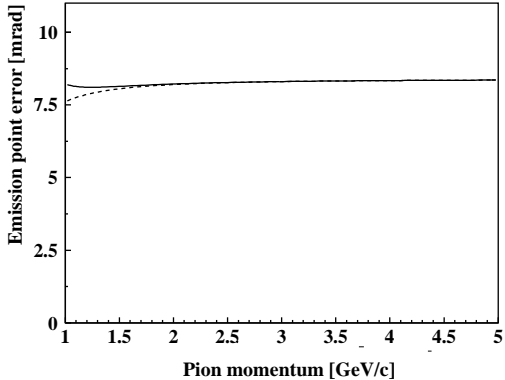}}
   \vskip -1cm
  \caption{\label{fig:focus-p}%
  Emission point error for a focusing combination (two 2~cm thick tiles with refractive indices 
1.05 and 1.042 at a distance of 18~cm from the photon detector) as a function of 
momentum for perpendicularly incident pions (full line) compared to the 
single radiator of half the thickness (dashed).
  }
 \end{center}
\end{figure}

The optimization of the radiator parameters thus clearly depends on the kinematical region
the counter should cover. In the case of the upgrade of the Belle particle identification
system, the indices are optimized for
tracks with $p=4$~GeV/$c$, the highest particle momenta from two-body $B$ decays. As can be
seen from Fig.~\ref{fig:focus-p},  the overlap of the two rings 
remains sufficiently good even at lower momenta, so that the emission point error (full line)
does not  significantly increase  with respect to the single radiator case of half the thickness 
(dashed line).%???
\begin{figure}
 \begin{center}
%  \centerline{\includegraphics[scale=.5]{figures/ai_sigth_alpha_phi_mul.eps}}
  \centerline{\includegraphics[scale=.5]{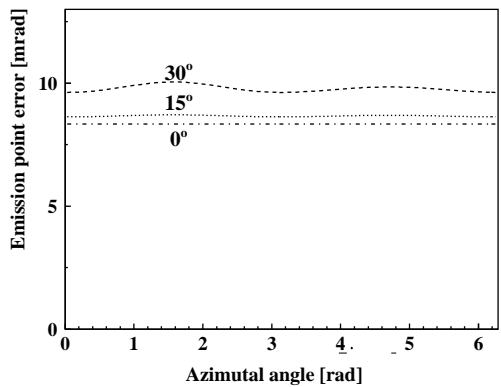}}
  \vskip -1cm
  \caption{\label{fig:focus-incang}%
  Emission point error for the focusing combination (same as in Fig.~\ref{fig:focus-p})
as a function of the  azimuthal angle around the track direction for 4~GeV/$c$ pions
 entering the counter at different angles of incidence ($0^\circ$ full line,
$15^\circ$ dash-dotted, $30^\circ$ dotted).
  }
 \end{center}
\end{figure}

Similarly, since the relation derived above is valid only for perpendicularly incident tracks,
 the performance has to be checked also as a function of incidence angle. As shown 
in Fig.~\ref{fig:focus-incang}, only a modest increase in the emission
point error is  expected up to 
 the maximal angle of incidence in the present application 
to the Belle upgrade (around 30$^\circ$). 
%Again, no degradation is expected  down to $p=1$~GeV/$c$.%???

For the dual radiator in the 
defocusing combination, the requirements are much less demanding. 
The two rings have to be well separated, and none of them should coincide with the 
ring corresponding to the other particle species. If one  requires
 for the case of $\pi/K$ separation
 that the kaon ring from the upstream radiator (corresponding to the angle $\theta_1(K)$) 
 is separated  by  $k \sigma_{\theta}$
 from the pion ring from the downstream radiator tile 
(corresponding to the angle $\theta_2(\pi)$), 
one arrives at the following expression
 $ (1+d/L) \tan{\theta_1(K)} -  \tan{\theta_2(\pi)} = k \sigma_{\theta}/\cos^2{\theta_2(\pi)}$,
from which one may derive a relation between $n_1$ and $n_2$.
For a specific example with $d/L=0.1$, $n_1=1.05$, $p=4$~GeV/$c$ and 
$\sigma_{\theta}=14$~mrad,  values typical for such a counter, 
we get $n_1-n_2 = 0.031$ for $k=10$.

Note that for both the focusing and defocusing combinations the performance of the
counter can be further optimized by varying the ratio of the thicknesses of the two
radiators and the total thickness of the radiator.  

\section{Experimental Apparatus}

\begin{figure}[t]
 \begin{center}
%  \centerline{\includegraphics[scale=.6]{figures/exp-setup2.eps}}
  \centerline{\includegraphics[scale=.6]{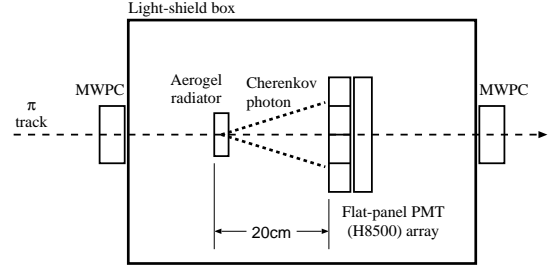}}
  \vskip -1cm
  \caption{\label{fig:exp-setup}%
  The experimental set-up. 
  }
 \end{center}
\end{figure}

We have performed two beam tests in March and June 2004.
The tests were carried out at the KEK-PS $\pi 2$
and T1 beam lines, where pions with momenta up to $4~\GeV/c$ are available.
The experimental set-up shown in Fig.~\ref{fig:exp-setup}
is basically the same as in the previous beam test~\cite{Matsumoto:2003wv}.
The counter is composed of one or more layers of aerogel radiator
and a photon-detection plane, parallel to the radiator face at a
distance of $20~\mathrm{cm}$.

Multi-anode PMTs (Hamamatsu H8500) were used as photo-detectors.
A total of 16 PMTs were positioned in a $4 \times 4$ array and aligned at
a $52.5~\mathrm{mm}$ pitch.
The surface of each PMT is divided into 64 ($8 \times 8$) channels
with $6.0~\mathrm{mm} \times 6.0~\mathrm{mm}$ pixel size.
This type of PMT is not immune to the magnetic field
and cannot be applied in the Belle spectrometer,
so  this device is considered as an intermediate step in our development.
For the final design, we plan to use the Hamamatsu HAPD or the BURLE
Micro-Channel Plate PMT~\cite{burle}.

As radiators the same set of aerogel samples was used as in the 
tests with single refractive index radiators~\cite{Matsumoto:2003wv}. 
These aerogels have indices between $1.01$ and $1.05$; the 
 transmission lengths  at 400~nm ($\Lambda$) are  around $30-40~\mathrm{mm}$.
In addition,  newly produced aerogels with refractive indices
up to $1.07$ were used, as well as two-layer aerogel samples~\cite{konishi}, 
where a single tile is comprised of two layers with different indices.

\section{Measurements and results}

\begin{figure}[t]
 \begin{center}
%  \centerline{\includegraphics[scale=.6]{figures/ai_march-focus-ring.eps}}
  \centerline{\includegraphics[scale=.6]{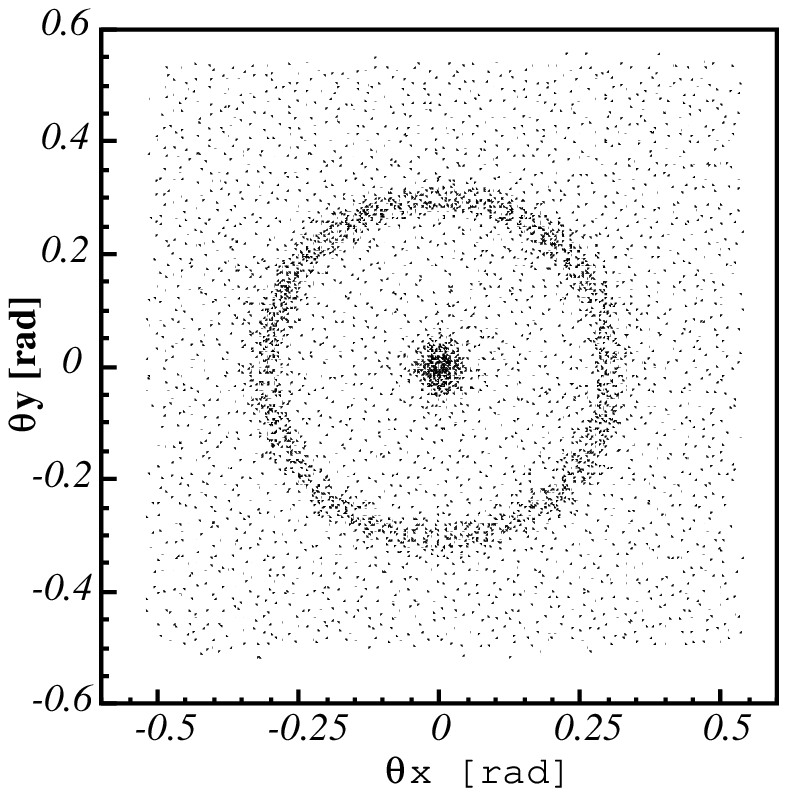}}
  \vskip -1cm
  \caption{\label{fig:march-focus-ring}%
  A typical distribution of PMT hits in the Cherenkov $x$, $y$ space with two 
 $20~\mathrm{mm}$ thick radiators ($n = 1.047$ and $n = 1.057$) in the focusing combination.
  }
 \end{center}
\end{figure}

A typical distribution of accumulated hits  on the photon detector
is shown in Fig.~\ref{fig:march-focus-ring}.
Cherenkov photons from the two  aerogel radiators are clearly
seen as a single ring with a low background level.
The hits near the center of the ring are due to Cherenkov
radiation generated by the beam particle in the PMT window.

\subsection{Test results with dual radiators}

\begin{figure}[t]
 \begin{center}
%  \centerline{\includegraphics[scale=.75]{figures/ai_march-focus-angle.eps}} 
  \centerline{\includegraphics[scale=.75]{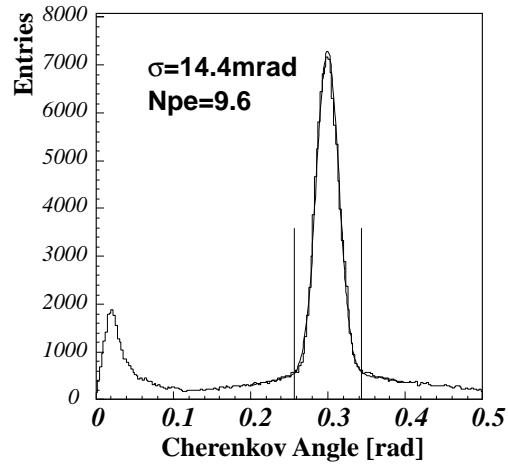}} 
  \vskip -1cm
  \caption{\label{fig:march-focus-angle}%
  Distribution of the Cherenkov angle for accumulated single photons
   from $4~\GeV/c$ pions in the case of a focusing dual radiator combination
   with $20~\mathrm{mm}$ thick radiators ($n = 1.047$ and $n = 1.057$).}
 \end{center}
\end{figure}

We first tested the  focusing dual radiator combination with
aerogel tiles of $n = 1.047$, $\Lambda = 34~\mathrm{mm}$
in the upstream position,
and $n = 1.057$, $\Lambda = 25~\mathrm{mm}$
in the downstream position.
Both radiators have a thickness of $20~\mathrm{mm}$.
Figure~\ref{fig:march-focus-angle} shows the resulting distribution
of the Cherenkov angle for single photons.  Note that the Cherenkov angle is 
calculated by
assuming that the photon is emitted in the middle of the combined radiator.
The basic parameters of the counter, the resolution and the number of
detected photons, are obtained by fitting to this distribution a
Gaussian function for the signal and a second order polynomial
for the background.  The resulting single photon
resolution $\sigma_{\theta}$ is $14.4~\mrad$, while the average number 
detected photons amounts to $\Npe = 9.6$. The Cherenkov angle resolution 
per track is calculated to be $4.8~\mrad$,
corresponding to a $4.8\sigma$ $\pi/K$ separation at $4~\GeV/c$.

For comparison, measurements were also performed for single radiator
cases when only the upstream or the downstream aerogel tile
 of the dual radiator is used.
%For the  upstream radiator, the yield was $6.9$ photons per ring and the single 
%photon resolution  $13.8$~mrad, while for the downstream radiator the two values 
%were $7.5$ and $17.2$~mrad.
For the  upstream and downstream radiator, the measured yield was 
$6.9$ and  $7.5$ photons per ring, respectively. The single photon 
resolution  was $13.8$~mrad for the single radiator in  the upstream position. 

The single photon resolution $\sigma_\theta=14.4~\mrad$ for the dual radiator 
is similar to that for the single radiator of half the thickness,
while the number of detected photons is larger for the dual radiator.
In the dual radiator RICH, a fraction of photons emitted in the upstream radiator
is scattered in the downstream aerogel.
Therefore, the photoelectron yield is expected to be
$\Npe \sim N_1 \exp(- d_2/(\cos\theta_1\Lambda_2)) + N_2 = 10.6$,
where $N_1$ ($N_2$) is the number of the detected photons
in the single radiator RICH with only upstream (downstream) radiator,  $\theta_1$
is the Cherenkov angle in the upstream radiator,
while $d_2$ and $\Lambda_2$ are the thickness and transmission length
of the downstream radiator.
This is consistent with the measured value of $9.6$.

\begin{figure}[t]
 \begin{center}
%   \centerline{\includegraphics[scale=.75]{figures/march-focus-nim-180-u.eps}}
   \centerline{\includegraphics[scale=.75]{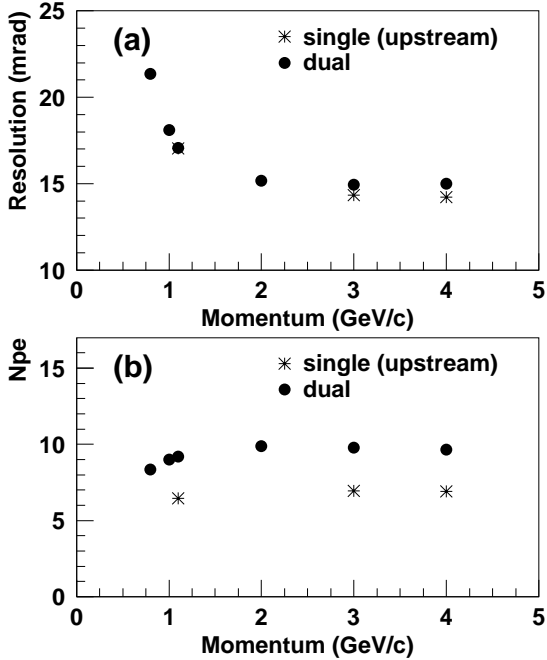}}
  \vskip -1cm
  \caption{\label{fig:march-focus}%
  Momentum dependence of the
   single photon resolution (a) and  the number of detected photons (b)
  for the dual radiator focusing combination compared to the single 
radiator.
  }
 \end{center}
\end{figure}

We also measured the momentum dependence of
$\Npe$ and $\sigma_\theta$ as shown in Fig.~\ref{fig:march-focus}.
Note that although the overlap of the rings from two radiators is 
optimized for a certain momentum,
the variation of  $\sigma_\theta$ is for  the dual radiator RICH
 similar to that of
the single radiator RICH over the full kinematic range. 
The increase in $\sigma_\theta$ at lower momenta is
consistent with the contribution from multiple scattering.

\begin{figure}[t]
 \begin{center}
% \centerline{\includegraphics[scale=.75]{figures/ai_march-defocus-angle.eps}} 
  \centerline{\includegraphics[scale=.75]{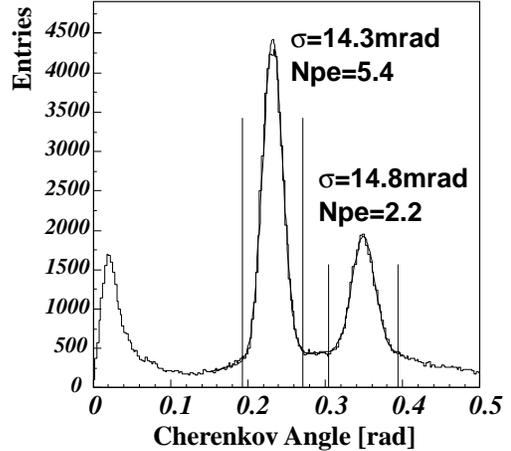}} 
  \vskip -1cm
  \caption{\label{fig:march-defocus-angle}%
  Distribution of the Cherenkov angle of single photons from $4~\GeV/c$ pions
  for a defocusing dual radiator RICH with $n_1=1.057$ and $n_2=1.027$.
  }
 \end{center}
\end{figure}

For the defocusing dual radiator set-up, we used
a combination of radiators with $n=1.057$ and $n=1.027$.
Each radiator had a thickness of $20~\mathrm{mm}$.
The single photon
Cherenkov angle distribution for this combination is shown
in Fig.~\ref{fig:march-defocus-angle}.
 Two well separated rings are clearly seen.
The sum of the number of photons from the two rings is larger than
that from a single radiator  with a thickness of 
$20~\mathrm{mm}$, while the resolutions are almost the same.
% Hence, the defocusing dual radiator RICH is also expected to have
% better performance than the single radiator RICH.
Note that unfortunately  the transparency of the downstream
radiator  was low ($\Lambda=19$~mm) in this particular measurement; 
the use of a typical sample
of $n \approx 1.03$ with the transmission length around 30~mm would considerably
increase the light yield at the outer ring.

We also tested a two-layer aerogel tile with indices of $1.060$
and $1.030$ in the defocusing combination. We could observe two well 
separated rings on the photon detector.
However, the photon yield was rather low because of a 
short transmission length of the radiator. This is
due to the production procedure for such kind of tiles, which 
 has not been optimized yet.

\subsection{Tests with  multiple radiators}

In order to study the multiple radiator RICH in the focusing combination,
we prepared a dual radiator configuration with  indices of 1.046 and 1.051,
and a triple radiator configuration with indices of 1.046, 1.051 and 1.056.
The thickness of each radiator was $10~\mathrm{mm}$, such that
 the dual radiator is $20~\mathrm{mm}$ thick, and
the triple radiator is $30~\mathrm{mm}$ thick.
The performance of these set-ups is compared with the single radiator 
combination with $n=1.046$ and
radiator thickness of $10~\mathrm{mm}$, $20~\mathrm{mm}$
and $30~\mathrm{mm}$.

\begin{figure}
 \begin{center}
%   \centerline{\includegraphics[scale=.7]{figures/ai_june-focus-nim.eps}}
    \centerline{\includegraphics[scale=.7]{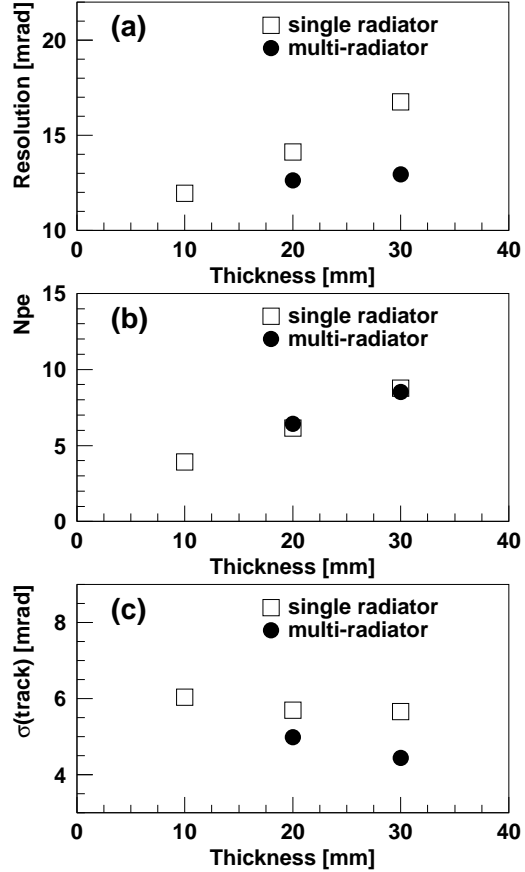}}
 \vskip -1cm
  \caption{\label{fig:june-focus}%
  Single photon resolution (a), number of detected photons (b)
  and single track resolution (c)
  for the focusing multiple radiator RICH combinations.
  The multiple radiator with a $20~\mathrm{mm}$ thick radiator is a dual
 radiator,   while that with $30~\mathrm{mm}$ is a triple radiator.
  }
 \end{center}
\end{figure}

Figure~\ref{fig:june-focus} shows $\sigma_\theta$, $\Npe$
and $\sigmatrack$ distributions against the thickness
of the radiators.
$\Npe$ increases as the radiator becomes thicker,
and there is no significant difference of $\Npe$
between single and multiple radiator combinations.
On the other hand, $\sigma_\theta$ of the single radiator RICH
increases considerably as the radiator gets thicker,
but $\sigma_\theta$ for the multiple radiator case increases only slightly.
As a result, $\sigmatrack$ is improved
by introducing the multiple radiator combination,
and the triple radiator with $30~\mathrm{mm}$ gives the best $\sigmatrack$
of 4.5~mrad.

We also tested a defocusing multiple radiator  with 
four radiator indices of $1.051$, $1.056$, $1.029$ and $1.034$
in a combination shown in Fig.~\ref{fig:multi-radiator}.e.
We compared it with a defocusing dual radiator RICH
with radiators of $n = 1.056$ and $1.025$.
The total thickness of aerogel radiators was $40~\mathrm{mm}$ in 
both cases.
The measured values of $\sigma_\theta$ and $\Npe$ are listed
in Table~\ref{tab:june-defocus}. As expected,
we find that the multiple radiator RICH has a better overall %??
angular resolution at roughly the same photon yield.

\begin{table}
 \begin{center}
  \caption{\label{tab:june-defocus}%
  Performance of the defocusing combination: single photon angular 
  resolution and photon yield
  for the multiple radiator and dual radiator combinations.}
  \begin{tabular}{lcc}
   \hline\hline
   RICH & Inner ring & Outer ring \\ \hline
   Dual   & $\sigma_\theta = 15.0~\mathrm{mrad}$
   & $\sigma_\theta = 14.6~\mathrm{mrad}$ \\
    radiator & $\Npe = 4.6$ & $\Npe = 1.7$ \\ \hline
   Multiple & $\sigma_\theta = 13.7~\mathrm{mrad}$
    & $\sigma_\theta = 13.0~\mathrm{mrad}$ \\
    radiator & $\Npe = 3.8$ & $\Npe = 2.1$ \\ \hline \hline
  \end{tabular}
 \end{center}
\end{table}

\section{Conclusion}

We have studied dual and multiple radiator combinations of
the proximity focusing aerogel RICH detector.
For the focusing combination,
we confirmed that the number of detected photons
could be increased without deteriorating angular resolution 
compared to the single radiator RICH.
In the triple radiator configuration we have achieved a 
Cherenkov angle resolution per track of $\sigmatrack = 4.5~\mrad$, which
corresponds to a $5.1\sigma$ $K/\pi$ separation
at $4~\GeV/c$.

In the defocusing dual radiator RICH, we observed
two rings with good angular resolution that were well separated.
We also succeeded to improve the resolution of each ring
with a RICH with four radiators.

Although the dual and multiple radiator proximity focusing 
RICH looks promising both
in the focusing and defocusing combination,
further studies are needed before deciding
which of the two combinations should be employed, and
 how many radiators should be used in the final design
for the upgraded Belle detector.
The method of reconstruction and the effect of background
will be studied in detail, especially for the defocusing combination
where two rings are produced.%?? 
We also plan to test the multiple radiator RICH with even more layers,
and investigate the production of multi-layer aerogel tiles with 
a large transmission length.

\section*{Acknowledgment}
We would like to thank the Belle collaboration
for their assistance in this project.
We would also like to thank Matsushita Electric Works Ltd.,
Hamamatsu Photonics K.K.\ and Meisei Co.\ for a fruitful
collaboration in the development of the aerogel, PMT and readout system.
This work was supported in part by a Grant-in-Aid for Scientific
Research from the Ministry of Education, Science
and Culture of Japan under Grant No.~13640306, 14046223, 14046224 and 16028208,
and in part by the Ministry of Higher Education,
Science and Technology of the Republic of Slovenia.


\begin{thebibliography}{99}
 \bibitem{bib:superkekb-loi}
	 K. Abe \textit{et al.} (edited by S.~Hashimoto, M.~Hazumi, J.~Haba,
	 J.~W.~Flanagan and Y.~Ohnishi), ``Letter of Intent for KEK Super B
	 Factory'', KEK report 2004-04, http://belle.kek.jp/superb/.
 \bibitem{Mori:2000cg} A.~Abashian et al. (Belle Collaboration),
	 Nucl. Instr. and Meth. \textbf{A479} (2002) 117.
 \bibitem{KEKB:NIM}
	 S.~Kurokawa and E.~Kikutani,
	 Nucl. Instr. and Meth., \textbf{A499}, 1 (2003),
	 and other papers included in this Volume.
 \bibitem{Iijima:2000uv} T.~Iijima et al., Nucl. Instr. and Meth.
	 \textbf{A453} (2000) 321.
 \bibitem{Matsumoto:2003wv} T.~Matsumoto et al., Nucl. Instr. and Meth.
	 \textbf{A521} (2004) 367.
  \bibitem{focus-first}  P.~Kri\v zan, ``Aerogel RICH'', talk given at
Super B Factory Workshop In Hawaii, 19-22 Jan 2004, Honolulu, Hawaii,
http://www.phys.hawaii.edu/superb04; S.~Nishida et al.,
	 ``Studies of a Proximity Focusing Aerogel RICH for the Belle Upgrade'', 
to be published in Proceedings of the IEEE Nuclear 
Science Symposium, Rome, Italy, October 17-22, 2004.
 \bibitem{burle} S.~Korpar et al.,
	 ``Tests of the BURLE 85011 Micro-Channel Plate PMT as the Detector
	 of Cherenkov Photons'', to be published in Proceedings of the IEEE Nuclear 
Science Symposium, Rome, Italy, October 17-22, 2004.
\bibitem{konishi} M.~Konishi et al.,
	 ``Development of New Silica Aerogel for the RICH Radiator
	 of the Super Belle Detector'', 
 to be published in Proceedings of the IEEE Nuclear Science Symposium, Rome, Italy, October 17-22, 2004.
\end{thebibliography}
\end{document}